\documentstyle[emulateapj,epsf,12pt]{article}
\begin{document}

\title{The MAHOROBA Project --- Deep Survey with an 
Optical Intermediate-Band Filter System on the Subaru Telescope}

\author{Yoshiaki Taniguchi}

\affil{Astronomical Institute, Graduate School of Science,
        Tohoku University, Aramaki, Aoba, Sendai 980-8578, Japan}

\begin{abstract}
We present a summary of the new optical intermediate-band
filter system for the prime-focus camera, Suprime-Cam, on the 
Subaru telescope at Mauna Kea Observatories. We also discuss a future
plan to promote a new deep survey with this filter system (the 
MAHOROBA project).
\end{abstract}

\section{Introduction}

Recent great progress in the observational astronomy
have revealed that a large number of high-redshift galaxies
can be accessible by continuum emission of galaxies
(stellar continuum, thermal continuum from dust grains,
or nonthermal continuum from plasma heated by supernovae)
in a wide range of observed wavelengths between optical and radio
(e.g., Williams et al. 1996, 2000; Lanzetta et al. 1996; Madau et al. 1996;
Barger et al. 1998; Hughes et al. 1998).

Another important technique to probe high-$z$ young galaxies is to
search for strong emission-line (e.g., Ly$\alpha$) galaxies.
It has been often argued that
forming galaxies at high redshifts experienced very luminous
starbursts and thus they could be much brighter in line emission such as
Ly$\alpha$ and [O {\sc ii}]$\lambda$3727 emission lines.
However, the low-resolution spectroscopic capability of 8-m class
telescopes is accessible only for galaxies with a magnitude
brighter than $B \sim$ 24.5 mag (e.g., Steidel et al. 1996a, 1996b).
Therefore,  even now, it is difficult to investigate
properties of very faint galaxies with $B > 25$, most of
which may be interesting high-$z$ galaxies.

In order to improve our understanding of the nature of such faint galaxies,
we would like to promote a new optical deep surveys with intermediate-band
filters whose spectroscopic resolutions are significantly higher
than those of typical broad-band filters.
In this article,
we present a short summary of the new optical intermediate-band
filter system for the prime-focus camera, Suprime-Cam
(Miyazaki et al. 1998),  on the
Subaru telescope (Kaifu 1998; Kaifu et al. 1998)
at Mauna Kea Observatories. We also discuss a future
plan to promote a new deep survey with this filter system (the 
MAHOROBA project\footnote{MAHOROBA is an ancient Japanese word. The 
meaning of this word is ^^ ^^ the best place" or ^^ ^^ the most 
comfortable place". Since the Subaru telescope must be a MAHOROBA
for Japanese optical/infrared astronomers, we call our new deep survey
program MAHOROBA.}).

\section{The Filter System}

The Subaru intermediate-band filter (IBF) system consists of 22 filters with
a spectral resolution of
$R = \lambda/\Delta\lambda \approx 23$,
covering a wavelength range between 3830 \AA~ and 9900 \AA~ (Table 1;
see also Hayashino et al. 2000).
The final specifications of these filters are summarized in Table 2.
Their nice performance
was already demonstrated by the wonderful image of the Ring Nebula M57
(Komiyama et al. 2000; see also Nature 401, 314).
In Figure 1, we show the transmission curves for all the IBFs.
The merits of our IBF system are summarized below.

I) {\it More reliable estimate of photometric redshifts of very faint galaxies}:
The accuracy of photometric redshifts using our IBF system is $\sim$90\%
(Shioya et al. 2001),
being much higher than those using 
broad-band photometric information, i.e., $\sim$60\% (e.g., Hogg et al. 1998).
Therefore, our filter system enables us to investigate physical properties of
faint galaxies down to 27AB and thus contribute to the understanding of 
cosmic star formation history as well as the large scale structure
at high redshift.

II) {\it More accurate estimate of SEDs of  high-$z$ galaxies}:
Since the spectroscopic resolution of IBF system is $R=23$,
we will be able to  investigate spectral energy distribution (SEDs)
of high-$z$ faint galaxies more accurately.
Our data are also useful in investigating the SED of old ($\sim 10^9$
years old) galaxies at intermediate redshift, e.g., $z \sim 1$
(e.g., Dunlop et al. 1996; Cowie et al. 2001).
Therefore, deep surveys with the present IBF system
allow us to investigate the cosmic star formation history
from high-$z$ through intermediate-$z$ to the present day unprecedentedly.

III) {\it More efficient capability of searching for strong emission-line objects}:
Deep surveys with a narrow-band filter provide a powerful method to
search for strong emission-line galaxies. Indeed, $\sim$ 100
high-$z$ Ly$\alpha$ emitters have been found recently
(Cowie \& Hu 1998; Keel et al. 1999; Steidel et al. 2000).
However, it is expected from the above recent surveys that
a huge number of such strong emission-line sources have not yet been probed
by the existing deep broad-band surveys because the majority of them have
too faint continuum emission fluxes and thus they cannot be seen in
broad-band images (see also Hu et al. 1999).

For Ly$\alpha$ emitters whose continuum magnitudes are brighter than
27 AB (see section 3), we will be able to detect them surely by
examining the continuum depression shorter than
$\lambda < \lambda_{{\rm Ly}\alpha}$, which is an important
characteristic of  high-$z$ Ly$\alpha$ emitters.
Even for Ly$\alpha$ emitters whose continuum magnitudes are fainter than
27 AB, we will be able to detect them surely if their emission-line
equivalent widths (EW) in the observed frame are larger than 200 \AA.

It is here remembered that the existing narrow-band deep surveys were made
for small selected volumes at high redshifts (e.g.,
$\sim 10^3 h^{-3}$ Mpc$^3$ where $h$ is defined with a Hubble constant of
$H_0 = 100 h$ km s$^{-1}$ Mpc$^{-1}$) and thus there
is no systematic search for strong emission-line objects of a significantly
large volume of the young universe. Thanks to the wide-field coverage of
Suprime-Cam (34$^\prime \times 34^\prime$), our survey can probe a volume
of more than $10^6 h^{-3}$ Mpc$^3$ in a field.

Strong emission-line galaxies could be much more common at high redshift.
If this is the case, it seems dangerous to
investigate the cosmic star formation history solely using galaxies found
in broad-band deep surveys, such as Lyman break galaxies (LBGs).
This also reinforces the importance of deep surveys with our IBF system.

In summary, deep and wide-field imaging with the combination between
this IBF system and Suprime-Cam on Subaru
will contribute very much to the understanding of cosmic star formation
history in the universe
and the growth of large-scale structures in the universe.


\begin{deluxetable}{ccccc}
\small
\tablenum{1}
\tablecaption{The Subaru IBF System \label{tbl-1}}
\tablewidth{0pt}
\tablehead{
\colhead{IBF} &
\colhead{Name} &
\colhead{Central Wavelength} &
\colhead{Band Width} &
\colhead{$z$(Ly$\alpha$)} \\
\colhead{No.} &
\colhead{} &
\colhead{( ${\rm \AA}$ )} &
\colhead{( ${\rm \AA}$ )} &
\colhead{}
}
\startdata
1 & IBF392 & 3922 & 193 & 2.22 \nl
2 & IBF409 & 4091 & 201 & 2.36 \nl
3 & IBF427 & 4267 & 207 & 2.51 \nl
4 & IBF445 & 4451 & 220 & 2.66 \nl
5 & IBF464 & 4643 & 230 & 2.82 \nl
6 & IBF484 & 4844 & 240 & 2.98 \nl
7 & IBF505 & 5053 & 250 & 3.16 \nl
8 & IBF527 & 5271 & 260 & 3.33 \nl
9 & IBF550 & 5499 & 270 & 3.52 \nl
10 & IBF574 & 5736 & 280 & 3.72 \nl
11 & IBF598 & 5984 & 295 & 3.92 \nl
12 & IBF624 & 6242 & 310 & 4.13 \nl
13 & IBF651 & 6512 & 325 & 4.36 \nl
14 & IBF679 & 6793 & 340 & 4.59 \nl
15 & IBF709 & 7086 & 340 & 4.83 \nl
16 & IBF738 & 7381 & 340 & 5.07 \nl
17 & IBF768 & 7676 & 340 & 5.31 \nl
18 & IBF797 & 7971 & 340 & 5.56 \nl
19 & IBF827 & 8266 & 340 & 5.80 \nl
20 & IBF856 & 8561 & 340 & 6.04 \nl
21 & IBF907 & 9070 & 410 & 6.46 \nl
22 & IBF965 & 9650 & 500 & 6.94 \nl
\enddata
\end{deluxetable}



\begin{deluxetable}{lll}
\small
\tablenum{2}
\tablecaption{The Final Specifications
              for the IBF System \label{tbl-2}}
\tablewidth{0pt}
\tablehead{
 \colhead{Item} &
 \colhead{} &
 \colhead{}
}
\startdata
Clear aperture     &      & 185 mm $\times$ 150 mm  \nl
Peak transmittance ($T_{\rm peak}$) & & $>$ 70\% ($>$ 80 \% goal) \nl
Homogeneity of $T_{\rm peak}$   &      & $<$ 5 \%                \nl
Ripple (valley/peak)     &      & $>$ 85\%                \nl
Linear change (valley/peak) &      & $>$ 90\%             \nl
CWL tolerance      &      & $< \pm $0.25 \% of CW         \nl
FWHM tolerance     &      & $< \pm $0.25 \% of CW         \nl
Bubble & d $<$ 0.1 mm     & acceptable              \nl
       & d $=$ 0.1-0.2 mm & $\leq$ 5 bubbles        \nl
       & d $=$ 0.2-0.5 mm & $\leq$ 3 bubbles        \nl
       & d $>$ 0.5 mm     & Not allowed             \nl
Stain              &      & Not allowed             \\
\enddata
\end{deluxetable}

\begin{figure*}
\epsscale{1.5}
\plotone{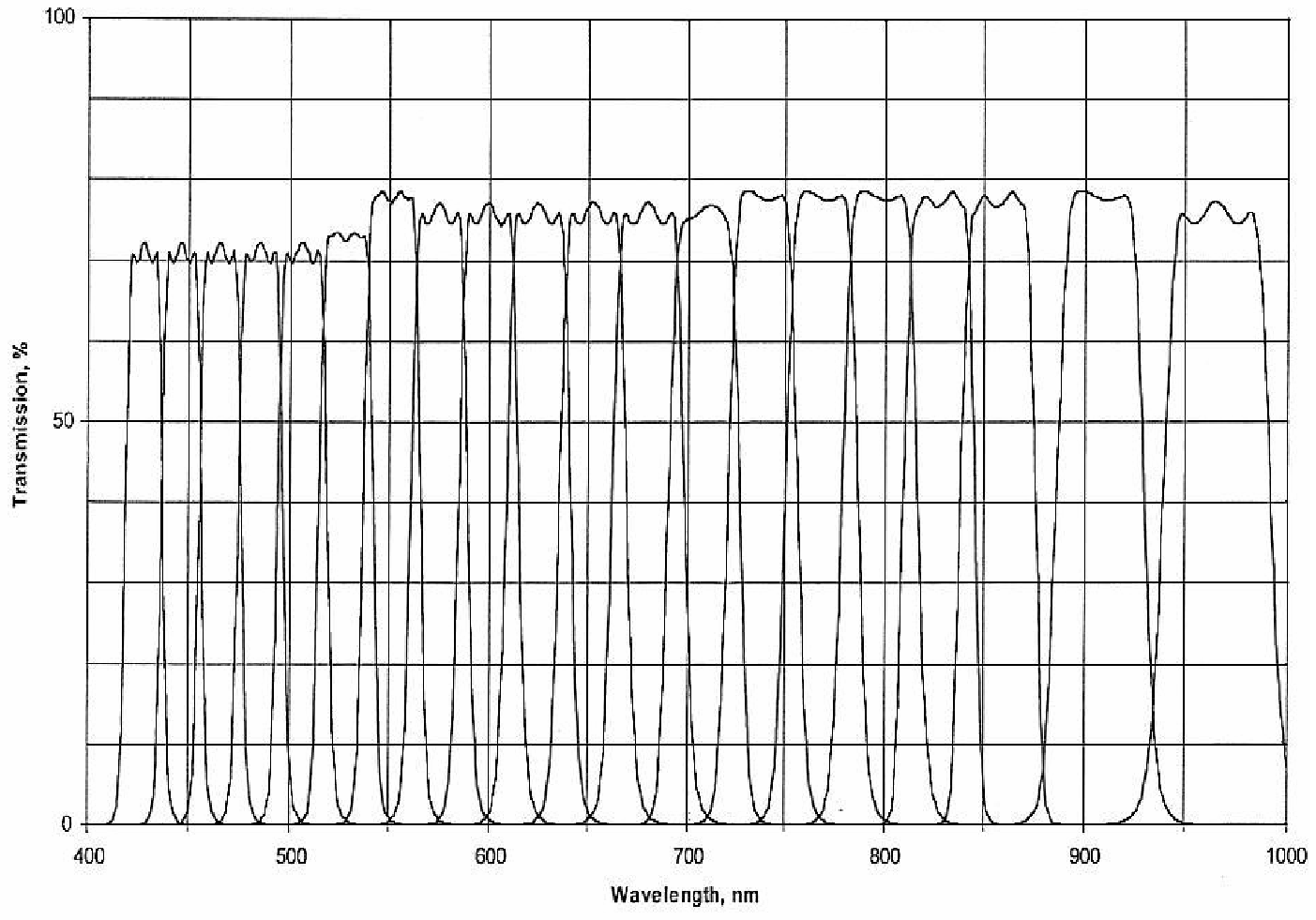}
\caption[]{
Transmission curves for the Subaru intermediate-band filters.
Note that the transmission curves of the first two filters are
not drawn in this Figure.
\label{fig1}
}
\end{figure*}

\section{The MAHOROBA Project}

The 8-m class telescope facilities such as the W. M. Keck  telescopes,
VLTs, and Subaru have the greatest capability of
probing high-$z$ universe.
Among the 8-m class telescopes, Subaru has a very unique merit;
i.e., the wide-field prime-focus camera, Suprime-Cam, whose 
sky coverage is 34$^\prime \times 34^\prime$. This instrument enables
us to perform wide-field, deep imaging surveys in the optical.

However, in the 90's, the Hubble Deep Field project (Williams et al. 1996)
already brought us the very deep images of the universe down to 29AB.
Even though its field is not so wide, we have learned so many things
from this survey together with the HDF-S project (Williams et al. 2000).
Therefore, simple deep imaging surveys with typical broad-band filters
on the 8-m class telescopes will not give us a new impact even if
we use Supreme-Cam on Subaru. We then believe that a new frontier
will be high-spectral-resolution, deep imaging surveys. This is the main
reason why we are now planning to promote a
new deep survey program called ^^ ^^ MAHOROBA" using the present 
IBF system on Subaru. 

In this project,
we will observe some blank sky fields such the Hubble Deep Field-North
(Williams et al. 1996) and the Subaru Deep Field (Maihara et al. 
2001). The limiting magnitude of this survey is tentatively set
to be 27AB for each IBF. 
Our main aims are; i) to investigate the origin of 
reionization of the universe, and ii) to search for primeval (metal-free) 
galaxies, because there has been no such an attempt in the optical.

According to the big bang cosmology, our universe began 12-14 billion 
years ago. The universe was filled with hot plasma for the first 
300,000 years (corresponding to redshift 
$z \sim$1000) and then the plasma recombined after that epoch. 
However, it is known that the universe (the intergalactic space)
between redshift 0 and 5 is completely ionized. 
Therefore, our universe was reionized by
some ionization sources between $z \sim$1000 and $z \sim$5
(i.e., the dark age: Rees 1996, 1999; see also Loeb \& Barakana 2001). 
The origin of reionization of the universe has been in debate
but there is no firm answer even at present. 
Since galaxies could form under the ultraviolet background
radiation responsible for the reionization, it is very 
important to investigate the origin of reionization.

However, observational properties of high-$z$ intergalactic space have 
been traditionally investigated by using quasar
absorption-line systems as well as the so-called 
Gunn-Peterson test. Therefore, any information of this kind
is indirect because neutral gas clouds are used in such studies. 
The MAHOROBA survey will be able to detect many faint emission-line
objects between $z \approx$2 and 7 and thus to investigate what sources
are mainly responsible for the reionization of universe
unambiguously.

It seems also worth noting that two very extended Ly$\alpha$ emission-line
nebulae, Ly$\alpha$ blobs (LABs), have been found by Steidel et al. (2000);
see also Francis et al. (2001).
Their observational properties are summarized as below
(we adopt an Einstein-de Sitter cosmology with
a Hubble constant $H_0 = 100 h$ km s$^{-1}$ Mpc$^{-1}$);
1) the observed Ly$\alpha$ luminosities are $\sim 10^{43} h^{-2}$
ergs s$^{-1}$, 2) they appear elongated morphologically,
3) their sizes amount to $\sim$ 100 $h^{-1}$ kpc,
4) the observed line widths amount to $\sim 1000$ km s$^{-1}$, and
5) they are not associated with strong radio-continuum sources
such as powerful radio galaxies.
As for the origin of LABs, two alternative ideas have been proposed.
One is that these LABs are
superwinds driven by the initial starburst in galaxies because all the above
properties as well as the observed frequency of LABs can be explained
in terms of the superwind model (Taniguchi \& Shioya 2000).
The other idea is that LABs are cooling radiation from proto-galaxies
or dark matter halos (Haiman, Spaans, \& Quataert 2000; Fardal et al. 2001;
Fabian et al. 1986; Hu 1992).
Standard cold dark matter models
predict that a large number of dark matter halos collapse at high redshift
and they can emit significant Ly$\alpha$ fluxes through collisional excitation
of hydrogen. These Ly$\alpha$ emitting halos are also consistent with
the observed linear sizes, velocity widths, and Ly$\alpha$ fluxes of the LABs.
Very recently, one of these Ly$\alpha$ blobs has
been detected at submillimeter wavelengths, 450 $\mu$M and 850 $\mu$m
(Chapman et al. 2001). Its rest-frame
spectral energy distribution between optical and far-infrared
is quite similar to that of Arp 220, which is a typical ultraluminous
starburst/superwind galaxy in the local universe. Therefore,
it is strongly suggested that the superwind model proposed by
Taniguchi \& Shioya is applicable to this Ly$\alpha$ blob
(Taniguchi, Shioya, \& Kakazu 2001).
Since the blob is more luminous in the infrared by a factor of 30 than Arp 220,
it comprises a new population of hyperwind galaxies at high redshift.
It is expected that such unknown objects will be found
in our survey.

We human being have not yet found very young,
primeval galaxies which do not contain heavy elements.
This suggests that any material was once polluted
by so-called Population III objects formed at $z \sim$ 10 -- 15.
Actually we know that broad emission-line
gas clouds associated with high-z quasars ($z>5$) are already
polluted chemically (Hamann \& Ferland 1999). 
However, a question arises here; ^^ ^^ have we
already made very deep surveys dedicated
to finding such primeval galaxies ?".
We think that the answer is ^^ ^^ no" because there has been no efficient tool
for this purpose; note that we need wide-field,
very deep imaging surveys with narrower-band filters.
It is known that metal-free, population III stars have higher effective
temperatures and thus they show very
strong He{\sc ii}$\lambda$1640 emission line without any metallic 
lines such as C{\sc iv}, and so on (e.g., Tumlinson, Girouz, \& Shull 2001
and references therein). This means that searches
for strong He{\sc ii} emission-line sources may lead to the discovery 
of primeval galaxies. Our survey will be also dedicated to this issue.

This project will be an 
unprecedented trial toward the dark age from the
side of the optical astronomy. 
Now we are ready to go the realm of high-redshift blizzard.

\vspace{0.5cm}

The author would like to thank the organizers of this Japanese-German
workshop, Nobuo Arimoto and Wolfgang Duschl for their warm encouragement. 
He would also like to thank many nice colleagues who have been working
together on the IBF system presented in this article, Keiichi Kodaira, 
Sadanori Okamura, Maki Sekiguchi, Mamoru Doi, Kazuhiro Shimasaku,
Satoshi Miyazaki, Yutaka Komiyama, Masafumi Yagi, Masami Ouchi,
Tomoki Hayashino, Yasuhiro Shioya, Takashi Murayama, \& Tohru Nagao.


\end{document}